\documentclass[journal]{IEEEtran}
\usepackage{graphicx}
\graphicspath{ {./plots/} }
\usepackage{amsmath}
\usepackage{color,soul}
\usepackage{subcaption}
\usepackage{caption}
\usepackage{multirow}
\usepackage{amsmath}
\usepackage{epstopdf}
\usepackage{xcolor}
\usepackage{booktabs}
\definecolor{tulga}{rgb}{1,0,0}
\definecolor{James}{rgb}{0,1,0}

\title{Online terrain estimation for autonomous vehicles on deformable terrains}
\author{James Dallas,
		Kshitij Jain, Zheng Dong, Michael P. Cole, Paramsothy Jayakumar, and
		Tulga Ersal$^*$%
		\thanks{This work was funded by U.S. Department of Defense under the prime contract number W56HZV-17-C-0005.}
		\thanks{J. Dallas, K. Jain, Z. Dong, and T. Ersal are with the Department of Mechanical Engineering, University of Michigan, Ann Arbor, MI 48109.}
		\thanks{M.P. Cole and P. Jayakumar are with the U.S. Army Ground Vehicle Systems Center, Warren, MI 48092.}
		\thanks{* Corresponding author: tersal@umich.edu}
		\thanks{DISTRIBUTION STATEMENT A. Approved for public release; distribution unlimited. OPSEC \#2439.}
}

\begin{document}
\bstctlcite{IEEEexample:BSTcontrol}
\maketitle

\begin{abstract}
	In this work, a terrain estimation framework is developed for autonomous vehicles operating on deformable terrains. Previous work in this area usually relies on steady state tire operation, linearized classical terramechanics models, or on computationally expensive algorithms that are not suitable for real-time estimation. To address these shortcomings, this work develops a reduced-order nonlinear terramechanics model as a surrogate of the Soil Contact Model (SCM) through extending a state-of-the-art Bekker model to account for additional dynamic effects. It is shown that this reduced-order surrogate model is able to accurately replicate the forces predicted by the SCM while reducing the computation cost by an order of magnitude. This surrogate model is then utilized in a unscented Kalman filter to estimate the sinkage exponent. Simulations suggest this parameter can be estimated within 4\% of its true value for clay and sandy loam terrains. It is also shown that utilizing this estimated parameter can reduce the prediction errors of the future vehicle states by orders of magnitude, which could assist with achieving more robust model-predictive autonomous navigation strategies.

\end{abstract}

\section{Introduction}
Autonomous ground vehicles (AGVs) have drawn interest for military applications to perform tasks, such as supply transport, in unsafe environments that could pose a threat to human operators \cite{Iagnemma2002}. Three considerations about military AGVs are important to motivate this work. First, military vehicles often need to operate off-road on deformable terrains, where the vehicle's mobility is dependent on the highly nonlinear tire forces generated at the tire-terrain interface \cite{Taheri2015}. Second, increasing the mobility of military AGVs is a critical need \cite{liu2017}. Third, state-of-the-art approaches to navigate such vehicles typically rely on model dependent architectures, such as Model Predictive Control (MPC) \cite{liu2017,Liu2018}. Therefore, when the AGVs are operated on deformable terrains, a more accurate knowledge of the terrain parameters becomes a critical enabler to maximize the mobility of the AGVs.

Much research has been performed in developing terramechanics models for off-road applications, which can be divided into empirical models, physics-based models, semi-empirical models  \cite{Taheri2015}.  Empirical model are the simplest; however, such models do not generalize well beyond the experimental test conditions used for their development. On the other hand, physics-based finite and discrete element models have proven to be of the highest fidelity, but the large computational efforts required renders them infeasible for real-time tire force prediction, thus limiting their applicability for use in AGVs and real-time terrain estimation \cite{Taheri2015}.  More promising candidates, and perhaps the most widely used, for real-time tire force prediction on deformable terrains are the semi-empirical models based upon the classical terramechanics theory developed by Bekker, including the Soil Contact Model \cite{Gallina2014,Ishigami2007,Smith2014,Guo2016}.  In these models, the tire is typically assumed rigid and the deformation is assumed to occur only in the terrain \cite{Smith2014}.  To model the complex tire-terrain interactions, these terramechanics models rely on knowledge of terrain-specific parameters such as cohesion, internal friction angle, or sinkage exponent.  During vehicle operation, these parameters may not be explicitly known or may be varying due to non-uniform terrains.  Therefore, real-time terrain estimation is necessary in AGVs to improve the accuracy of the terramechanics models online and generate better informed control commands. Having this capability would also provide insight into traversability of terrains, such that path planning algorithms can reroute the vehicle to avoid regions where loss of mobility or excessive power consumption is likely to occur \cite{Howard2006}.

Researchers have already recognized this need and a limited number of results are available in the literature. In particular, in \cite{Gallina2014,Gallina2016}, a Bayesian procedure is utilized for terrain parameter identification, but making this approach work online is subject to future research.  Other researchers have proposed an online algorithm for estimating soil cohesion and internal friction angle utilizing a linear least-squares estimator for a rover \cite{Iagnemma2002,Iagnemma2004}.  The algorithm relies on simplifying classical terramechanics equations through linear approximations to increase computational efficiency and subjects the rover to periodic high and low speed traverses \cite{Iagnemma2002}.  However, linear approximations can lead to inaccurate stress approximations \cite{ZhenzhongJia2011}, and hence inaccurate force prediction, and periodically operating at low speeds is not desirable when maximum mobility is desired. Hence, online estimation of deformable terrain parameters for off-road AGVs is still an open research area and is the focus of this work. 

This study presents a new approach for online terrain parameter estimation. First, due to the large computation time associated with integrating stresses in SCM and limitations of classical terramechanics equations, a nonlinear reduced-order model is developed by extending the work  presented in \cite{ZhenzhongJia2011} to account for additional dynamic effects such that a sufficient agreement with SCM can be achieved. Then, the reduced-order terramechanics model is incorporated in a 3 DoF bicycle model \cite{Liu2018} to create an estimation model, whereas the actual vehicle is represented with a 11 DoF plant model with SCM. The predictions from the estimation model are fused with measurements from the plant model in an Unscented Kalman Filter (UKF) to identify the dominant terrain parameter, namely, the sinkage exponent.  The result is an online terrain estimation approach that can be used to better inform control and path-planning algorithms for AGVs.

The rest of this paper is organized as follows. Sec. \ref{sec:TerramechanicsModels} first briefly reviews the SCM model used in the plant model. Then a state-of-the-art fast terramechanics model used as a benchmark is introduced and the significant deviations of its predictions from SCM are demonstrated. This model is then modified to improve its accuracy vis-\`a-vis SCM, so that a suitable estimation model is obtained. Sec. \ref{sec:VehicleModels} presents the vehicle models, both the plant model as well as the estimation model. The terrain estimation procedure based on UKF is summarized in Sec. \ref{sec:TerrainEstimation}. Sec. \ref{sec:Results} gives the simulation results including the accuracy of the estimations and their ability to improve the predictive accuracy of the 3 DoF model.  Finally, Sec. \ref{sec:Conclusion} gives the conclusions drawn from this work.

\section{Terramechanics Models} \label{sec:TerramechanicsModels}
\subsection{Soil Contact Model (SCM)}
This section briefly reviews the terramechanics model adopted in this work to represent the tire-terrain interactions in the plant simulations with high fidelity. This model is also used to evaluate the accuracy of the fast terramechanics models, including a state-of-the-art model and the surrogate model developed in this work. As such, this model serves as the ground truth for the purposes of this work.

The terramechanics model used in this study for generating the lateral tire forces acting on the vehicle is based on the Soil Contact Model (SCM) reported in \cite{Gallina2014,Krenn2011}.  Verification of the model can be found in \cite{Krenn2009}.  The SCM calculates relevant forces and torques acting on a 3 dimensional object in contact with a deformable terrain as summarized below.  

The SCM algorithm relies on a discretized mesh of the tire and terrain to search for contact points at the tire-terrain interface.  In the contact detection step, the vertices of the tire mesh are projected onto the nearest vertices of the terrain digital elevation map, effectively arranging the contact vertices in individual columns.  The sinkage at each vertex can be determined from the minima of each column, assuming the vertex location is a point of sinkage.  The effective contact width, $b$, can then be determined from the footprint's area and contour length \cite{Krenn2011}.  

Following contact detection, the algorithm calculates the stresses at each contact node of the footprint as follows.  The pressure, $\sigma$, is expressed as \cite{Bekker1962}
\begin{equation}
\sigma = (k_c/b+k_\phi)h^n \label{sig_scm}
\end{equation}
The shear stress, $\tau$, is expressed as \cite{Janosi1961}
\begin{equation}
\tau = \tau_\text{max}(1-e^{-j/k}) \label{tau_scm}
\end{equation}
with $\tau_\text{max}$ given as
\begin{equation}
\tau_\text{max}=(c+\sigma\tan\phi) \label{tau_m_scm}
\end{equation}

\begin{table}
	\begin{center}
	\caption{SCM terrain parameters.}
	\label{tab: 1}
	\begin{tabular}{ccc}
		\toprule
		\bf Parameter                   & \bf Symbol          & \bf Unit \\
%		\midrule
		Cohesive modulus            & $k_c$           &$ \text{N}/\text{m}^{n+1}$ \\
%		\midrule
		Frictional modulus          & $k_\phi$        &$\text{N}/\text{m}^{n+2}$ \\
%		\midrule
		Sinkage exponent            & $n$               & - \\
%		\midrule
		Shear deformation modulus   & $k$               & m \\
%		\midrule
		Cohesion                    & $c$               & Pa \\
%		\midrule
		Angle of internal friction  & $\phi$          & rad \\
		\bottomrule
	\end{tabular}
	\end{center}
\end{table}

In the above expressions $h$ is the sinkage, $b$ is the tire effective width, and $j$ is the shear deformation.  The remaining parameters are internal parameters characterizing the terrain as summarized in Table \ref{tab: 1}.  
The forces generated at the tire-terrain interface can then be given by integrating the stresses over the entire contact patch.  The above overview is a summary of \cite{Gallina2014}; a more complete discussion is given in \cite{Krenn2011}.

SCM is a rather complex model due to the discretizations and integrations involved and may thus not be suitable for real-time parameter identification purposes.  It has been shown that the accuracy of SCM is heavily influenced by the discretization resolution \cite{Krenn2011}.  Furthermore, several SCM operations are of $N^2$ complexity, where $N$ is the number of grid nodes \cite{Krenn2011}.  As an example, for a discretization of just 200 total nodes (100 per tire in a bicycle model), the time required by SCM can be around 20 ms \cite{Krenn2011}. Furthermore, for the UKF, the estimation method used in this work, 17 sigma points must be generated  as discussed in Sec. \ref{sec:TerrainEstimation}, each calling the terramechanics model twice (once per tire in the bicycle model).  Thus the total time spent calculating tire forces can be around 350 ms per a single UKF iteration. Finally, taking into account that many UKF iterations are needed to achieve estimation convergence, a UKF with SCM can be expected to take several minutes to converge, which is impractically long. Therefore, faster terramechanics models are needed.

\subsection{State-of-the-Art Fast Terramechanics Model}\label{sec:state-of-the-art-fast-terramechanics-model}
Much less computationally demanding solutions better suited for online estimation are given by Bekker-based models.  These models are again based on \eqref{sig_scm}-\eqref{tau_m_scm}; however, $\sigma$, $\tau$, $h$, and, $j$ are now replaced by functions of the angle of contact, $\theta$.  As such,  \eqref{sig_scm}-\eqref{tau_m_scm} are rewritten as:
\begin{equation}
\sigma(\theta) = (k_c/b+k_\phi)h(\theta)^n \label{sigma}
\end{equation}
\begin{equation}
\tau(\theta) = \tau_\text{max}(1-e^{-j(\theta)/k}) \label{tau}
\end{equation}
\begin{equation}
\tau_\text{max}=(c+\sigma(\theta)\tan\phi) \label{tau_max}
\end{equation}
where $h(\theta)$ is given by
\begin{equation}
	\label{h1}
	h(\theta)=\begin{cases} 
	r(\cos\theta-\cos\theta_\text{f}) & \theta_\text{m} \le \theta \le \theta_\text{f} \\
	r(\cos\theta_\text{e}-\cos\theta_\text{f}) & \theta_\text{r} \le \theta \le \theta_\text{m} \\
	\end{cases}
\end{equation}
with
\begin{equation}
\theta_\text{m} = (a_0+a_1s)\theta_\text{f} \label{theta_m}
\end{equation}
\begin{equation}
\theta_\text{f} = \cos^{-1}(1-h_f/r) \label{theta_f}
\end{equation}
\begin{equation}
\theta_\text{e}=\theta_\text{f}-(\theta-\theta_\text{r})(\theta_\text{f}-\theta_\text{m})/(\theta_\text{m}-\theta_\text{r})\label{theta_e}
\end{equation}
\begin{equation}
\theta_\text{r} = \cos^{-1}(1-\Lambda h/r) \label{theta_r}
\end{equation}
where $r$ is the radius of the tire; $\theta_\text{f}$ is the angle at which the front of the tire comes into contact with the terrain; $\theta_\text{m}$ is the location of maximum normal stress with $a_0$ and $a_1$ as terrain parameters typically taking on values of 0.4 and 0--0.3, respectively \cite{Wong1967}; $s$ is the longitudinal slip of the tire; $\theta_e$ is the equivalent front contact angle for angles less than $\theta_m$; $\theta_\text{r}$ is the angle at which the rear of the tire loses contact with the terrain; and $\Lambda$ is a property of the terrain characterizing the sinkage ratio.

Finally, $j(\theta)$ is given as 
\begin{equation}
	\label{j_pos}
	j(\theta)=\begin{cases}
	r[(\theta_\text{f}-\theta)-(1-s)(\sin\theta_\text{f}-\sin\theta)] & s \ge 0 \\
	r[(\theta_\text{f}-\theta)-(1/(1+s))(\sin\theta_\text{f}-\sin\theta)] & s < 0
	\end{cases}
\end{equation}

The maximum sinkage can be calculated in an iterative fashion by using the Newton-Raphson method as proposed in \cite{Guo2016} as follows.  The maximum sinkage is initialized as the static sinkage, which is based on the load on the tire $W$:
\begin{equation}
h_0 = \left[\frac{3W}{b(3-n)(k_c/b+k_\phi)\sqrt{2r}}\right]^{\frac{2}{2n+1}} \label{h0}
\end{equation}
However, due to dynamic effects, such as slippage, additional sinkage is induced.  To account for this, the reaction force is calculated as 
\begin{equation}
F_z = \int_{\theta_\text{r}}^{\theta_\text{f}}rb(\tau(\theta)\sin(\theta)+\sigma(\theta)\cos(\theta))d\theta \label{Fz}
\end{equation}
and a new sinkage is determined as 
\begin{equation}
h_0' = h_0-F_z(h_0)/F'_z(h_0) \label{h0_newt}
\end{equation}
The iterative procedure terminates when the calculated reaction force is within a specified tolerance of the normal force applied to the tire.   Once the maximum sinkage is determined, the lateral force $F_y$ can then be calculated in a similar fashion as in \cite{Ishigami2007, Guo2016}, i.e.,
\begin{equation}
F_y = \int_{\theta_\text{r}}^{\theta_\text{f}}rb\tau_y(\theta) \label{Fy}
\end{equation}
with $\tau_y(\theta)$ given as
\begin{equation}
\tau_y(\theta) = \tau_\text{max}(1-e^{-|j_y(\theta)|/k_y}) \label{tau_y}
\end{equation}
where
\begin{equation}
j_y(\theta) = r(1-s)(\theta_\text{f}-\theta)\tan\beta \label{jy}
\end{equation}
and $\beta$ is the side slip angle.

\begin{figure}
	\centering
	\begin{subfigure}{0.35\textwidth}
		\centering
		\includegraphics[scale=0.6]{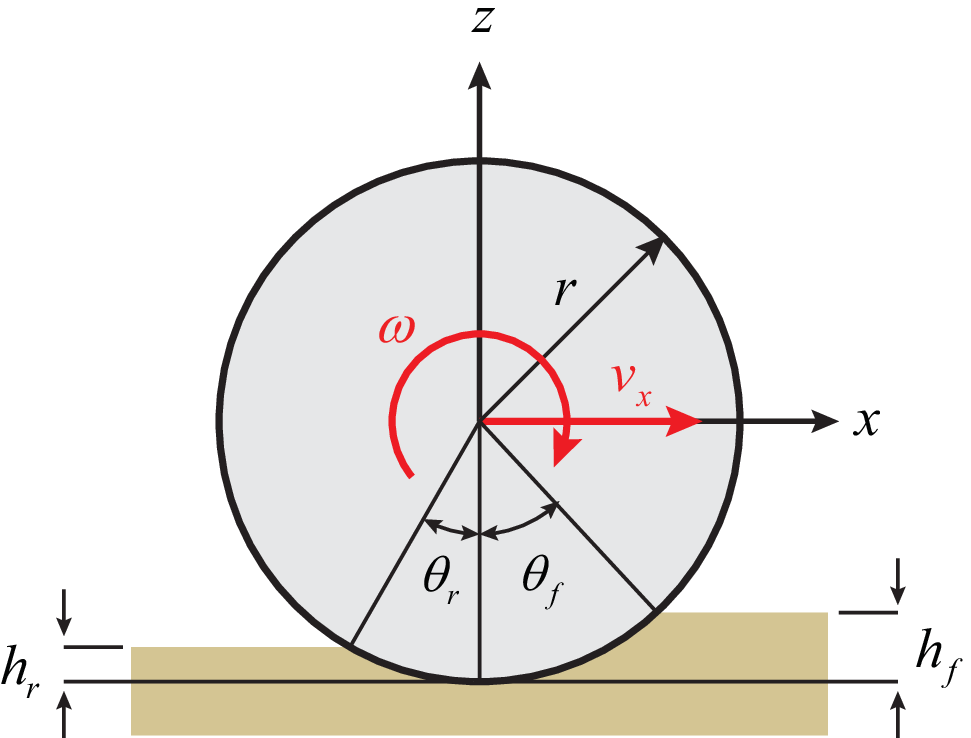}
		\subcaption{Side view}
	\end{subfigure}
	\begin{subfigure}{0.45\textwidth}
		\centering
		\includegraphics[scale=0.6]{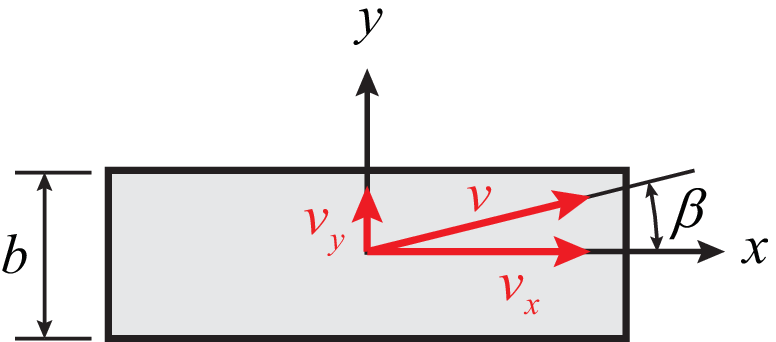}
		\subcaption{Top view}
	\end{subfigure}
	\caption{Tire-terrain geometry for positive slip.}%: (a) side view, (b) top view.}
	\label{fig:1}
\end{figure}

Depending on the soil type  $\tau_y(\theta)$ can also be represented with different a formulation such as
\begin{equation}
\tau_y(\theta) = \tau_\text{max}(j/k_y)(e^{1-j_y(\theta)/k_y}) \label{tau_y2}
\end{equation}
Other formulations can be found in \cite{Smith2014}. All relevant variables are depicted in Fig. \ref{fig:1}.

\begin{table}
	\begin{center}
		\caption{Terrain parameters for sand \cite{Guo2016}.}
		\label{tab: 2}
		\begin{tabular}{cc}
			\toprule
			\bf Parameter       & \bf Value\\
%			\hline
			$k_c$           &1000 ($\text{N}/\text{m}^{n+1}$)\\
%			\hline
			$k_\phi$        &1528600 ($\text{N}/\text{m}^{n+2}$)\\
%			\hline
			$n$               & 1.08 (--)\\
%			\hline
			$k$               & 0.024 (m)\\
%			\hline
			$c$               & 200 (Pa)\\
%			\hline
			$\phi$          & 0.4712 (rad)\\
			\bottomrule
		\end{tabular}
	\end{center}
\end{table}

\begin{table}
	\begin{center}
		\caption{Wheel states for benchmark simulation.}
		\label{tab: 3}
		\begin{tabular}{cc}
			\toprule
			\bf State       & \bf Value\\
%			\hline
			Normal load           &2500 (N)\\
%			\hline
			Longitudinal slip     &0.2 (--) \\
%			\hline
			Camber angle          & 0 (rad) \\
%			\hline
			Speed                 & 5.5 (m/s) \\
			\bottomrule
		\end{tabular}
	\end{center}
\end{table}

To assess the accuracy of \eqref{Fy} compared to SCM, a simulation is run in Chrono \cite{Chrono}.  The simulation utilizes a single wheel test bed operating on a sand-like terrain using Chrono's built-in SCM terrain. The test bed allows for individual control of the tire's velocity, load, longitudinal slip, and lateral slip. The terrain properties used in this simulation are representative of sand and given in Table \ref{tab: 2}.  The simulation sweeps the tire through a range of lateral slips with a 1 Hz sine wave.  The load, longitudinal slip, camber angle, and linear velocity of the tire are all held at the constant values given in Table \ref{tab: 3}.

\begin{figure}
	\centering
	\includegraphics[width=3in]{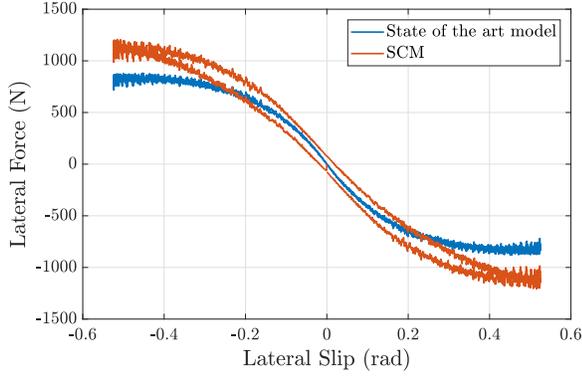}
	\caption{Simulation results for SCM (orange) and model based on \eqref{Fy} (blue).  The simulation uses the inputs given in Table \ref{tab: 3} and the terrain properties given in Table \ref{tab: 2}. }\label{fig:2}
\end{figure}

Fig. \ref{fig:2} shows the results of an SCM simulation run in Chrono (orange) and the force predicted by \eqref{Fy} (blue).  The term $k$ is assumed to be constant, rather than a function of lateral slip as in \cite{Ishigami2007}.  This is to maintain consistency with the SCM formulation used in Chrono.  As seen in the figure, the base model of \eqref{Fy} captures the overall trend, at least in the linear region around zero lateral slip, but averages the two distinct curves seen in SCM. This is because the current formulation does not account for the hysteresis effects of varying lateral slip; i.e., the shear deformation of \eqref{jy} does not account for the shearing resulting from the tire rotation that induces the lateral slip.  Note that in this work the lateral slip is varied by the steering angle applied to the tire.

Recognizing this shortcoming in the state-of-the-art fast terramechanics model, a new surrogate model for SCM is developed in the next section.

\subsection{New Surrogate Model} \label{surrogate}
This section presents the new terramechanics model developed as a fast surrogate for the SCM.

The new surrogate is obtained by replacing \eqref{tau_y} and \eqref{tau_y2} with the following expressions, respectively.
\begin{equation}
\tau^*_y(\theta) = \tau_\text{max}(1-e^{-|j^*_y(\theta)|/k^*_y}) g_1(v,s,F_z,n) \label{tau^*_y}
\end{equation}
\begin{equation}
\tau^*_y(\theta) = \tau_\text{max}(j/k^*_y)(e^{1-j^*_y(\theta)/k^*_y}) g_1(v,s,F_z,n) \label{tau^*_y2}
\end{equation}
with
\begin{align}
\begin{split}
j^*_y(\theta) = &-|r(1-s)(\theta_f-\theta)\tan\beta|+ \\
&\text{sign}(\beta)\big(r\sin(\theta) \Delta \delta\, g_2(v,s,F_z,n)\big) \label{jy_sur}
\end{split}
\end{align}
where $\Delta\delta$ is the step change in the steering angle. $k_y$, a parameter originally describing the shear displacement required to generate peak shear stress, is now empirically estimated as a function of the wheel states, i.e.,
\begin{equation}
	k^*_y = g_3(v,s,F_z,n)
\end{equation}

The lateral force acting on the vehicle is then determined as in \eqref{Fy}. It should be noted that an additional term in \eqref{Fy} is often given representing the bulldozing force; however, simulations suggest this contribution is minimal for this application.  Additionally, integrating the original nonlinear functions over the contact patch is a computationally demanding task. Therefore, the quadratic approximation proposed in \cite{ZhenzhongJia2011} is adopted in the surrogate model.  Furthermore, the modifications shown represent the lateral force acting on the vehicle frame, not the lateral forces in the tire reference frame.

Simulations covering the operating range of a notional military AGV are run to develop the modifying functions $g_1(\cdot)-g_3(\cdot)$. For each slip range of the clay simulation, as described in the Appendix, the simulations are run at 4 equispaced wheel loads, 5 slips, 5 translational velocities, and 7 sinkage exponents.  Other terrain parameters are set to their nominal values, because only the sinkage exponent is selected as the parameter to be estimated due to the higher sensitivity of tire forces to sinkage exponent than other parameters \cite{Gallina2014,Ishigami2008}. Table \ref{tab: 4} shows the range of inputs covered in the simulations.  
In these simulations, the inputs are held at constant values and the lateral slip is varied with a sinusoidal input.  Following this, correction factors are determined for $\tau_y$, $j_y(\theta)$, and $k_y$ to match the output of  \eqref{Fy} with each of the SCM simulations.  Least squares curve fitting is then used to derive the relationship between the correction factors and the simulation inputs of Table \ref{tab: 4}, resulting in the modification functions $g_1(\cdot)-g_3(\cdot)$. To ensure the model was not subject to overfitting, over 1,500 independent validation simulations were performed as described in the following paragraphs.  

\begin{table}
	\begin{center}
		\caption{Wheel states and terrain ranges for development of $g_1(\cdot)-g_3(\cdot)$}.
		\label{tab: 4}
		\begin{tabular}{cc}
			\toprule
			\bf State       & \bf Value\\
%			\hline
			Normal load           &1000--4000 (N)\\
%			\hline
			Longitudinal slip     &-0.9--0.9 (--)\\
%			\hline
			Camber angle          & 0 (rad) \\
%			\hline
			Speed                 & 2.5--8.5 (m/s) \\
%			\hline
			$n$                 &  0.4--1.3 (--)\\
			\bottomrule
		\end{tabular}
	\end{center}
\end{table}

\begin{figure}
	\centering
	\includegraphics[width=3in]{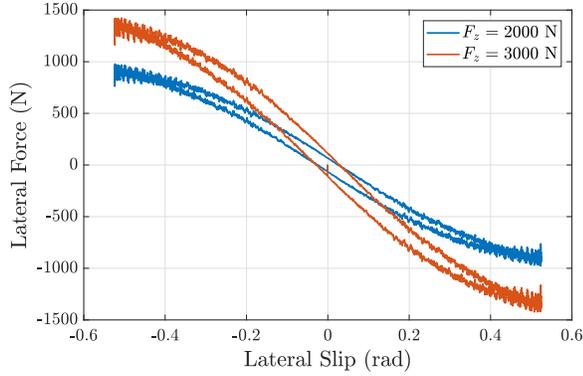}
	\caption{Simulation results for SCM with $F_z$ being 2000 N (blue) and 3000 N (orange).  The simulation uses the inputs given in Table \ref{tab: 3} and the terrain properties given in Table \ref{tab: 2}, except for the normal load and slip (-0.5).}\label{fig:Fz_eff}
\end{figure}

\begin{figure}
	\centering
	\includegraphics[width=3in]{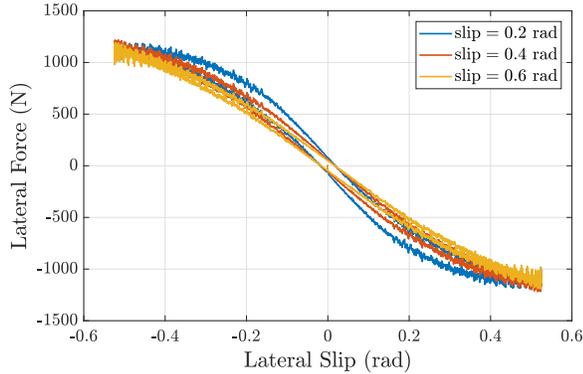}
	\caption{Simulation results for SCM with slip being 0.2 (blue), 0.4 (orange), 0.6 (yellow).  The simulation uses the inputs given in Table \ref{tab: 3} and the terrain properties given in Table \ref{tab: 2}, except for the slip.}\label{fig:slip_eff}
\end{figure}

\begin{figure}
	\centering
	\includegraphics[width=3in]{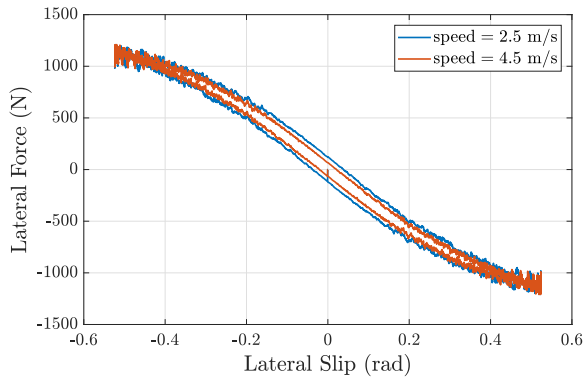}
	\caption{Simulation results for SCM with speed being 2.5 m/s (blue) and 4.5 m/s (orange).  The simulation uses the inputs given in Table \ref{tab: 3} and the terrain properties given in Table \ref{tab: 2}, except for the speed and slip (0.5).}\label{fig:spped_eff}
\end{figure}

Several parameters in the surrogate terramechanics model have distinct effects on the lateral force prediction and can be modified to achieve better agreement with SCM.  For illustration purposes, the effects of each input on a sand terrain are shown in Fig. \ref{fig:Fz_eff}-\ref{fig:spped_eff}.  The slope of the linear region can be set by modifying $k_y$ with $g_3(\cdot)$, the distance between the two curves can be set by adjusting $j_y$ with $g_2(\cdot)$, and the overall magnitude of the force can be adjusted with $g_1(\cdot)$.  The effect of the wheel states on  $g_1(\cdot)-g_3(\cdot)$ are as follows.  The effect of wheel load can be captured with linear functions for $g_1(\cdot)-g_3(\cdot)$.  Increased wheel load tends to increase the magnitude of the lateral force, increase the slope of the linear region, and increase the seperation between the top and bottom curves as shown in Fig. \ref{fig:Fz_eff}.  The effect of longitudinal slip can be modeled by polynomials for $g_1(\cdot)-g_3(\cdot)$. As seen in Fig. \ref{fig:slip_eff}, for positive slips, lower magnitude longitudinal slips tend to increase the slope of the linear region, while also causing a larger spread between the top and bottom curve.  The effect of translational velocity can be captured with a power function for $g_2(\cdot)$ alone, because it has minimal effect on $g_1(\cdot)$ and $g_3(\cdot)$. Hence, $g_1(\cdot)$ and $g_3(\cdot)$ do not depend on translational velocity. As seen in Fig. \ref{fig:spped_eff}, increased translational velocity tends to have little effect on the slope of the linear region, but reduces the hysteresis. Example formulations of $g_1(\cdot)-g_3(\cdot)$ for a clay terrain are given in the Appendix. 

\begin{figure}
	\centering
	\includegraphics[width=3in]{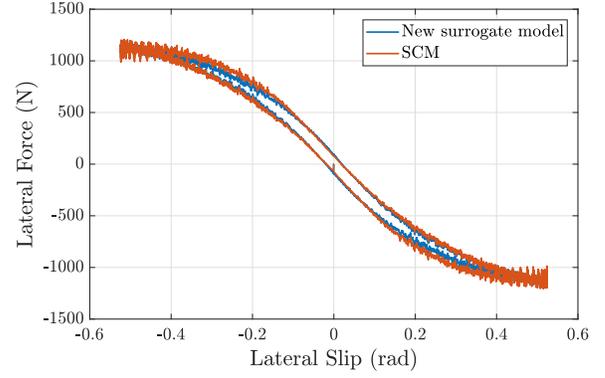}
	\caption{Simulation results for SCM (blue) and the new surrogate model (orange) for the inputs given in Table \ref{tab: 3} and the terrain properties given in Table \ref{tab: 2}.}\label{fig:3}
\end{figure}

Once $g_1(\cdot)-g_3(\cdot)$ are determined, over 1,500 
independent validation simulations are ran. The results of the surrogate model are shown in Fig. \ref{fig:3} (blue).   Much better agreement is observed between the surrogate model and the SCM simulation compared to Fig. \ref{fig:2}.  It should also be noted that the surrogate model runs in 200-400 $\mu$s, which is an order of magnitude more efficient than what is reported for SCM and more suitable for online terrain estimation.

\section{Vehicle Models} \label{sec:VehicleModels}
Two vehicle models are employed in this work; a 11 DoF model to represent the plant and a 3 DoF bicycle model to be used as part of the estimator. This section describes these models.
\subsection{Plant Model} \label{sec:PlantModel}

To represent the actual vehicle in the simulation-based validation of the proposed surrogate model and terrain estimator, the Chrono software is utilized to simulate the dynamics of a notional military vehicle as well as to implement the SCM \cite{Chrono}. For the purposes of this work, the vehicle is modeled with a double wishbone suspension, rack-pinion steering, 4 wheel drive, and a simple powertrain without a torque converter or transmission. This results in a 11 DoF vehicle model that is then combined with the SCM as the tire-terrain interaction model. The data received from the plant is then corrupted with Gaussian noise and serves as the measurement $y_k$ in \eqref{y_update} in Sec. \ref{sec:TerrainEstimation}.  Table \ref{tab: 5} lists the standard deviations used in the noise model for each state.  Actual sensors typically offer lower noise levels; hence the chosen standard deviations represent a worse-case scenario to test the ability of the estimator \cite{Ryu2002}.

\begin{table}
	\begin{center}
		\caption{Measurement standard deviations used for sensor simulation.}
		\label{tab: 5}
		\begin{tabular}{cc}
			\toprule
			\bf State       & \bf Noise ($\sigma$)\\
%			\hline
			$x$           &1.2 (m)\\
%			\hline
			$y$     &1.2 (m) \\
%			\hline
			$\psi$          & 0.0175 (rad) \\
%			\hline
			$u$                 & 0.25 (m/s) \\
%			\hline
			$v$               &  0.25 (m/s)\\
%			\hline
			$\omega_z$                 &0.0175 (rad/s)  \\
			\bottomrule
		\end{tabular}
	\end{center}
\end{table}

\subsection{Bicycle Model}
As part of the terrain estimation process that is detailed in Sec. \ref{sec:TerrainEstimation}, a vehicle model is needed to predict future vehicle states based on the tire forces from the surrogate model.  
For this work, a 3 DoF bicycle model with forward Euler integration is adopted, as it provides a proper level of fidelity while maintaining enough simplicity for short-horizon predictions \cite{Liu2016}.  The bicycle model takes on the following form:
\begin{equation}
\dot{z_b} = \begin{bmatrix}
u\cos\psi-(v+L_\text{f}\omega_z)\sin\psi \\
u\sin\psi+(v+L_\text{f}\omega_z)\cos\psi \\
w_z \\
a_x \\
(F_\text{yf}+F_\text{yr})/M_t-u\omega_z \\
(F_\text{yf}L_\text{f}-F_\text{yr}L_\text{r})/I_{zz}
\end{bmatrix} \label{bicycle}
\end{equation}
where the state vector, $z_b$, is defined as
\begin{equation}
z_b :=
\begin{bmatrix}
x \\ y \\ \psi \\ u \\ v \\ \omega_z
\end{bmatrix}
=
\begin{bmatrix}
\textrm{global $x$  position  of  front  axle}\\
\textrm{global  $y$  position  of  front axle}\\
\textrm{yaw  angle}\\
\textrm{longitudinal  velocity} \\ 
\textrm{lateral  velocity }\\ 
\textrm{yaw  rate}
\end{bmatrix} \label{bicycle_states}
\end{equation}
with $M_t$ being the vehicle mass, $I_{zz}$ being the vehicle's yaw moment of inertia, and $L_\text{f}$ and $L_\text{r}$ being the distance from the vehicle's center of gravity to the front and rear axles, respectively.  Finally, $F_\text{yf}$ and $F_\text{yr}$ are the lateral forces generated from the front and rear tires acting on the vehicle body, as obtained from the terramechanics model.

\section{Terrain Estimation} \label{sec:TerrainEstimation} 

The terrain parameter to be estimated is chosen as the sinkage exponent $n$, because it has been shown to be the dominant parameter \cite{Gallina2014}. All other terrain parameters are assumed to be some nominal values based on the specific terrain type, which can be determined from terrain classification algorithms such as the ones described in \cite{Howard2006,Weiss2008}.

To estimate the unknown terrain parameter $n$, it is appended to the 3 DoF bicycle model in \eqref{bicycle} with trivial dynamics. Here $n$ is given as a 2x1 vector to account for the front and rear tires.  This is to mitigate the influence of unmodeled multipass effects and in the case of a discrete terrain change where the front tire and rear tire may operate on different terrains.  The augmented state vector and state dynamics are given as
\begin{equation}
z :=\begin{bmatrix}
z_b  \\ n
\end{bmatrix}, \quad
\dot{z} =\begin{bmatrix}
\dot{z_b}  \\ 0
\end{bmatrix} \label{bicycle_est}
\end{equation}

Given the measurements of the vehicle states in \eqref{bicycle_states}, the augmented dynamics are utilized in an unscented Kalman filter (UKF) to estimate the augmented state vector in \eqref{bicycle_est} including the sinkage exponent. It is worth noting that many other algorithms are available in the literature for nonlinear parameter estimation, including, but not limited to,  extended Kalman filters, transitional Markov Chain Monte Carlo algorithms, and particle filters. Among these options the UKF is preferred in this work, because preliminary explorations suggest that the UKF offers a good balance between accuracy and computational speed for this application. 

The UKF is composed of two general steps; a time update step and a measurement update step.  Assume that a system is given in discrete time as:
\begin{equation}
z_{k+1} = F(z_k,v_k) 
\label{x_update}
\end{equation}
\begin{equation}
y_k = H(z_k,n_k) \label{y_update}
\end{equation}
where $z$ is the state, $y$ is the observation, and $v$ and $n$ are the process and observation noise, respectively.  The functions $F(\cdot)$ and $H(\cdot)$ are nonlinear functions describing the dynamics and outputs. In this application, $z$ takes the form of the state vector in \eqref{bicycle_est} and $F(\cdot)$ is obtained by discretizing the state equation in \eqref{bicycle_est} using the forward Euler method. $H(\cdot)$ is given as the state vector in \eqref{bicycle_states}.

First, a set of $2L+1$ sigma points are created to capture the statistical distribution of the states, where $L$ is the dimension of the state vector $z$. The sigma points are determined as follows:
\begin{equation}
Z_{k-1} = [\hat{z}\quad \hat{z} \pm  (\sqrt{(L+\lambda)P_z})_i] \label{sigma_pts}
\end{equation}
where $\hat{z}$ is the mean value of $z$. $\lambda$ is a scaling parameter given as:
\begin{equation}
\lambda = \alpha^2(L+\kappa)-L \label{lambda}
\end{equation}
where $\alpha$ is a tunable scaling parameter that typically takes a value between 0 and 1, $\kappa$ is another scaling parameter that is typically set to 0.  At the time update step, the sigma points are propagated through the original nonlinear system as:
\begin{equation}
Z^z_{k|k-1} = F(Z^z_{k-1},Z^v_{k-1}) \label{x_propigate}
\end{equation}
The following weights are then calculated
\begin{equation}
W^m_0 = \lambda/(L+\lambda) \label{W_0m}
\end{equation}
\begin{equation}
W^c_0 = \lambda/(L+\lambda) + (1-\alpha^2 + \zeta) \label{W_0c}
\end{equation}
\begin{equation}
W^{m,c}_i = 1/(2(L+\lambda)) \label{Wi}
\end{equation}
where $\zeta$ is set to 2 for Gaussian distributions. The statistics of the time update step are then given by:
\begin{equation}
\hat{z}^-_k = \sum_{i=0}^{2L}W^m_iZ^z_{i,k|k-1} \label{x_k-}
\end{equation}
\begin{equation}
P^-_k = \sum_{i=0}^{2L}W^c_i(Z^z_{i,k|k-1}-\hat{z}^-_k)(Z^z_{i,k|k-1}-\hat{z}^-_k)^T  \label{P_k-}
\end{equation}
\begin{equation}
Y_{k|k-1} = H(Z^z_{k|k-1},Z^n_{k-1}) \label{Y_k}
\end{equation}
\begin{equation}
\hat{y}^-_k = \sum_{i=0}^{2L}W^m_iY_{i,k|k-1}  \label{y_k-}
\end{equation}

Finally, the measurement update step is given by the following set of equations:
\begin{equation}
P_{\hat{y}_k\hat{y}_k} = \sum_{i=0}^{2L}W^c_i(Y_{i,k|k-1}-\hat{y}^-_k)(Y_{i,k|k-1}-\hat{y}^-_k)^T \label{Pyy}
\end{equation} 
\begin{equation}
P_{{z_k}y_k} = \sum_{i=0}^{2L}W^c_i(Z_{i,k|k-1}-\hat{z}^-_k)(Y_{i,k|k-1}-\hat{y}^-_k)^T  \label{Pxy}
\end{equation}
\begin{equation}
K = P_{{z_k}{y}_k}P_{\hat{y}_k\hat{y}_k}^{-1} \label{K}
\end{equation}
\begin{equation}
\hat{z}_k = \hat{z}^-_k +K(y_k-\hat{y}^-_k) \label{x_pred}
\end{equation}
\begin{equation}
P_k = P^-_k - KP_{\hat{y}_k\hat{y}_k}K^T \label{P_k}
\end{equation}
The above process is a summary of the algorithm given in \cite{Wan}.  Intuitively, the process works by merging model-based predictions of the states with their measurements from the plant by exploiting the uncertainties associated with each to determine the best estimates of the states.  For further discussion of UKF and details of its implementation, the reader is referred to \cite{Wan,Kolas2009}. It also worth noting that while this work only focuses on estimating the dominant parameter $n$, other terrain parameters could be estimated simultaneously, as well. However, this would incur additional computational costs as the state space dimension increases, thus increasing the number of sigma points necessary in the UKF.

\section{Results and Discussion} \label{sec:Results}

\begin{table}
	\begin{center}
		\caption{Terrain parameters for simulated terrains \cite{Smith2014}.}
		\label{tab: 6}
		\begin{tabular}{ccc}
			\toprule
			\bf Parameter       &\bf Sandy Loam	&\bf Clay\\
%			\hline
			$k_c$           &5300 ($\text{N}/\text{m}^{n+1}$) & 13200 ($\text{N}/\text{m}^{n+1}$)\\
%			\hline
			$k_\phi$        &1515000 ($\text{N}/\text{m}^{n+2}$)		&692200 ($\text{N}/\text{m}^{n+2}$)\\
%			\hline
			$n$               &0.7 (--) &0.5 (--)\\
%			\hline
			$k$               &0.025 (m)&0.01 (m)\\
%			\hline
			$c$               &1700 (Pa)& 4140 (Pa)\\
%			\hline
			$\phi$          &0.5061 (rad)&0.2269 (rad)\\
			\bottomrule
		\end{tabular}
	\end{center}
\end{table}

In this section the performance of the terrain estimator is evaluated. The performance is evaluated from two different points of view: (1) the accuracy of the estimated sinkage exponent $n$, and (2) the accuracy of the predicted state trajectories of the vehicle . The former assesses the algorithm's ability to find the true sinkage exponent, whereas the latter assesses the utility of estimating the sinkage exponent in the larger picture of predicting the future states of the vehicle.  Note that if the assumed nominal values for the non-estimated parameters are not representative of the true terrain type, then the estimator may not necessarily converge to the true terrain parameter, as it will attempt to find a value that compensates for the errors in the non-estimated values and achieves the best prediction capability of the vehicle model. For the ultimate aim of more accurately predicting the future mobility capabilities of the vehicle, the second evaluation criterion is the more relevant one.

\begin{figure} 
	\centering
	\begin{subfigure}{2.5in}
		\centering
		\includegraphics[width=2.5in]{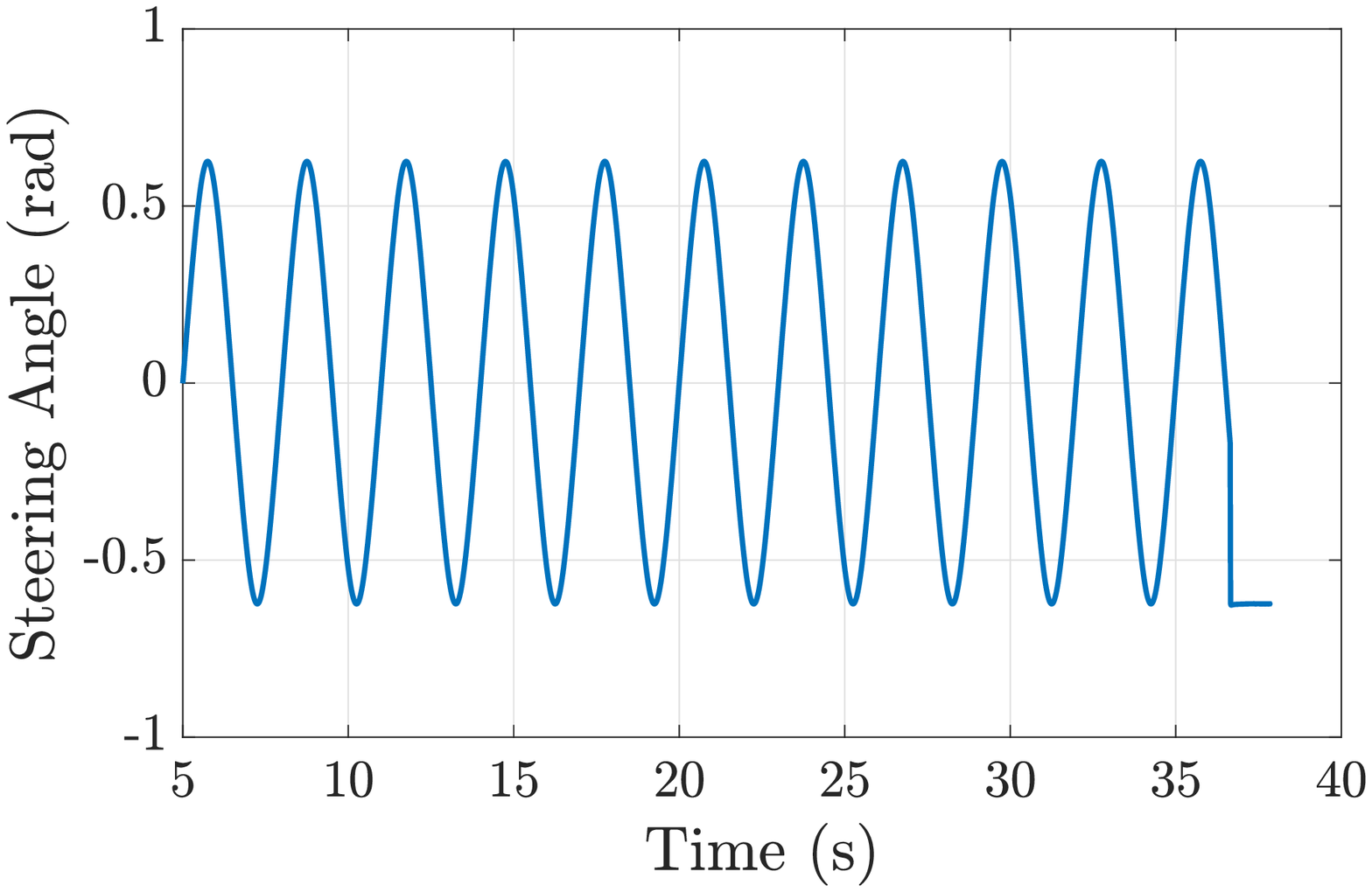}
		\subcaption{Steering}
	\end{subfigure}
	\begin{subfigure}{2.5in}
		\centering
		\includegraphics[width=2.5in]{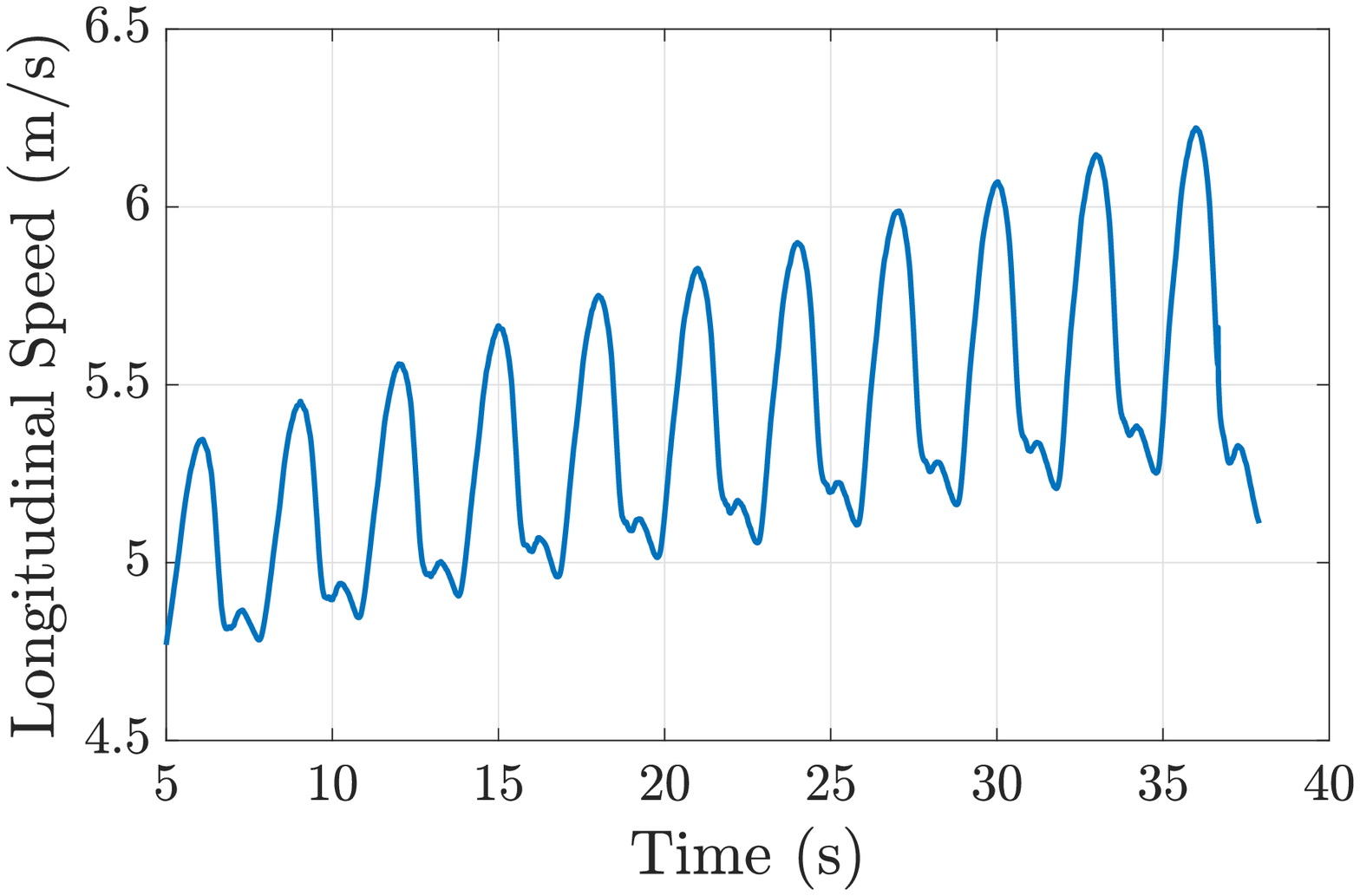}
		\subcaption{Longitudinal speed}
	\end{subfigure}
	\caption{Steering and velocity profiles used in simulation.}
	\label{fig: sim_profiles}
\end{figure}

Simulations are performed utilizing Chrono's SCM deformable terrain and the developed AGV model in Sec. \ref{sec:PlantModel}.  Two terrains are considered including sandy loam and clay.  Relevant terrain parameters are given in Table \ref{tab: 6}.  The simulation subjects the AGV to sinusoidal steering commands, steering fully to the left and right over a three second period.  The throttle is also varied with a sinusoidal command such that varying speeds are achieved. No braking command is given.  The applied steering and speed profiles for the clay simulation are shown in Fig. \ref{fig: sim_profiles}. Throttle and steering commands of the same frequency are given in the sandy loam simulation, as well. Two remarks are in order. First, as seen in Fig. \ref{fig: sim_profiles}, no requirement on constraining the vehicle to low speeds (on the order of 10 cm/s) is made, which is in contrast to previous efforts \cite{Iagnemma2004}.  This enables enhanced mobility, which is critical for military applications.  Second, a sinusoidal steering input is selected to induce lateral dynamics for the vehicle.  Since the bicycle model only utilizes the lateral forces acting on the vehicle, it is critical for the estimation that the vehicle operates in such a way that lateral dynamics are induced.  Otherwise, the lack of information on the lateral dynamics leads to parameter variations having negligible effects on the output of the bicycle model. In other words, if $F_\text{yf}$ and $F_\text{yr}$ are zero, it is not possible to estimate terrain parameters based on lateral forces.

In all simulations, the simulation time step is set to 2 ms in Chrono.  The purpose of these simulations is to determine the estimation algorithm's accuracy and utility under different terrain conditions.  Once the simulations are complete, noise is added to the outputs to simulate sensors, as discussed in Sec. \ref{sec:PlantModel}.  The estimator is then run at a 12 ms time step and the simulated measurements are received at every 24 ms.  The terrain parameter $n$ is initialized with a value off of the true terrain parameter used in the plant simulation.  The remaining terrain parameters are set to their true values given in Table \ref{tab: 6}. Note that the true values are used here only to assess how closely the algorithm can converge to the true sinkage exponent.

\begin{figure}
	\centering
	\begin{subfigure}{3in}
		\centering
		\includegraphics[width=2.5in]{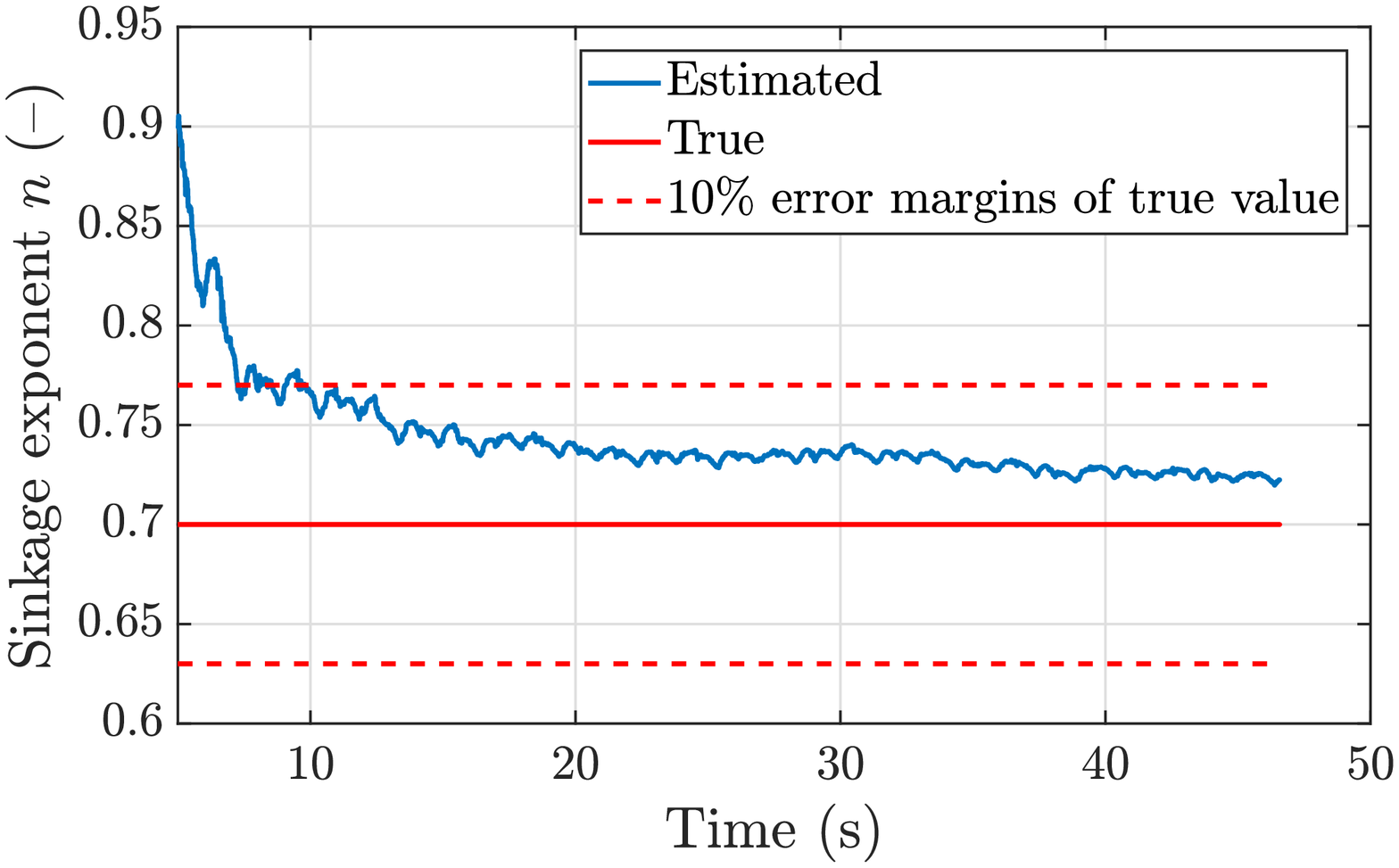}
		\subcaption{Sandy loam}
	\end{subfigure}
	\begin{subfigure}{3in}
		\centering
		\includegraphics[width=2.5in]{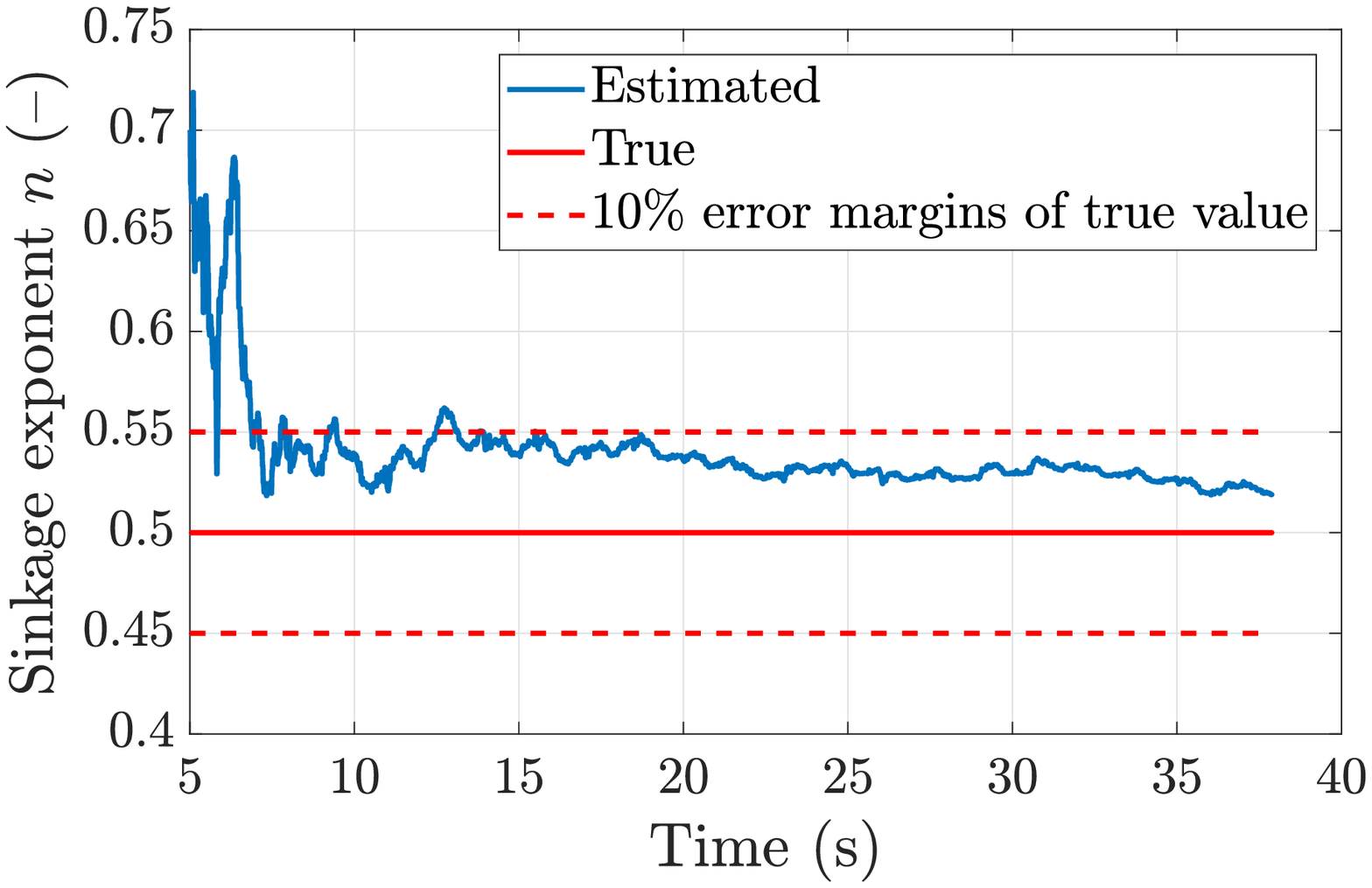}
		\subcaption{Clay}
	\end{subfigure}
	\caption{Simulated sinkage exponent estimation results.}\label{fig:4}
\end{figure}

Table \ref{tab: 7} displays the initial guess of the value of the sinkage exponent $n$, its converged estimate by the algorithm, and the error associated with the estimated parameter. The initial terrain parameter for sandy loam is chosen to be representative of Buchele (Michigan) sandy loam and the initial terrain parameter for clay is chosen to be representative of Thailand clay \cite{Wong2001}.   On both terrains, the percent error in the estimated terrain parameter is less than 4\%, where the estimated value is taken to be the final value by the end of the simulation. Fig. \ref{fig:4} shows the estimated terrain parameters for the two considered terrains as time evolves.   
The differences between the converged and true terrain values can be due to model discrepancies between the high fidelity Chrono model and the 3 DoF bicycle model along with discrepancies arising from the reduced order terramechanics model. Nevertheless, the estimator converges within 10\% of the estimated parameter within 5 seconds for both cases.

\begin{table}
	\begin{center}
		\caption{Initial guess, estimated value, and estimation errors of the sinkage exponent $n$ for simulated terrains. }
		\label{tab: 7}
		\begin{tabular}{ccccc}
			\toprule
			\bf Terrain       & \bf Initial guess 			&\bf True val.	&\bf Converged val. &\bf \% error\\
			%			\hline
			Sandy loam        &0.9 	&0.7 & 0.722  & 3.1\%\\
			%			\hline
			Clay               &0.7 	&0.5 & 0.519  & 3.8\%\\
			\bottomrule
		\end{tabular}
	\end{center}
\end{table}

The peak computation time of the estimator is 10.5 ms and 7.5 ms for clay and sandy loam simulations respectively, thus demonstrating the potential to achieve real-time estimation. The platform running this estimation consists of 16 GB Memory and a single core 3 GHz Intel Core i7 processor.

While estimating the terrain properties accurately is a worthy goal in and of itself, it is more of interest to evaluate to what extent the estimations can improve the predictive capability of the bicycle model as motivated above. 
To accomplish this second evaluation, the bicycle model, with the terramechanics model parameterized by either the initial guess or the converged terrain parameter, is used to predict the vehicle states approximately 0.5, 2.5, and 5.0 seconds into the future for the clay case.  After this time the vehicle states are reset to the true values received from Chrono. As such, this procedure mimics the operational procedure of a model predictive control approach, where a receding finite horizon optimal control problem is solved periodically with updated information available from sensors \cite{liu2017}.   

\begin{figure}
	\centering
	\includegraphics[width=3.5in]{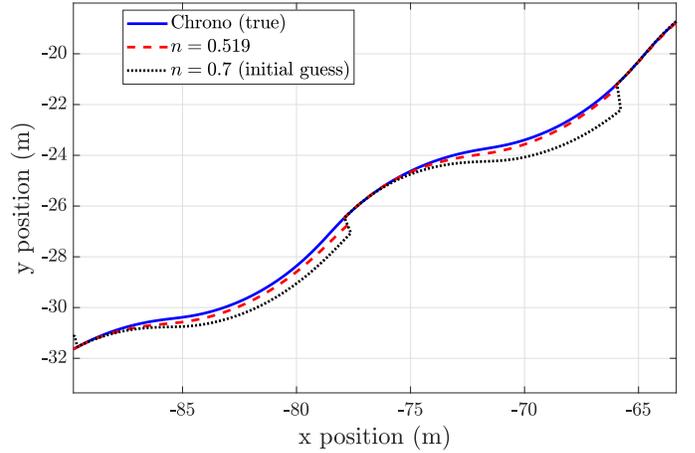}
	\caption{Simulated vehicle positions for AGV operating on clay.  True vehicle positions from Chrono (blue solid line), bicycle model parameterized by $n = 0.519$ (red dashed line), and bicycle model parameterized by initial terrain guess $n = 0.7$ (black dotted line). }\label{fig:5}
\end{figure}

Table \ref{tab: 8} depicts the mean squared errors (MSE) of the state estimates given by the 3 DoF bicycle model for both the case when the initial guess for the sinkage exponent for clay is used and the case when the converged estimate is used over the entire 32.89 s simulation. The model parameterized by the estimated terrain property yields significantly better predictions, especially at larger time horizons with order of magnitude reductions in MSE. Fig. \ref{fig:5} shows a portion of the simulation, using the $\sim$2.5s time horizon and depicting the true vehicle positions from the plant (blue solid line), and the predicted positions using the bicycle model with the initial guess of the sinkage exponent (black dotted line) and with the converged sinkage exponent (red dashed line). As can be seen, the converged value yields much more accurate predictions, thus demonstrating the ability of the estimator to significantly improve prediction fidelity. Similar results are also observed for the other terrains, but they are not reported here due to space limitations. This improvement in turn could lead to better performance in model predictive controllers, which is subject to future research.

\begin{table*}
	\begin{center}
		\caption{Mean squared error over entire simulation with varying prediction horizons for clay using estimated terrain parameter ($n = 0.519$) and initial guess ($n = 0.7$).}
		\label{tab: 8}
		\scriptsize
		\begin{tabular}{ccccccc}
			\toprule
			\bf Time horizon & \multicolumn{2}{c}{\bf 0.5 (s)} & \multicolumn{2}{c}{\bf 2.5 (s)} & \multicolumn{2}{c}{\bf 5 (s)} \\
			\midrule
			\bf State			&\bf n=0.519  &\bf n=0.7&\bf n=0.519  &\bf n=0.7&\bf n=0.519  &\bf n=0.7\\
			\midrule
			$x$           &0.0034 (m)&0.0025 (m)&0.037 (m)&0.01 (m)&0.075 (m)&0.078 (m)\\
			%			\hline
			$y$     &0.0035 (m) &0.0051 (m)&0.022 (m)&0.15 (m)&0.12 (m)&0.34 (m)\\
			%			\hline
			$\psi$          & 3.17e-05 (rad)& 1.8e-04 (rad)&2.45e-04 (rad)& 0.0089 (rad)&9.9e-04 (rad)& 0.0115 (rad) \\
			%			\hline
			$u$                 & 6.46e-05(m/s) & 6.46e-05 (m/s)& 1.33e-04 (m/s)& 1.33e-04 (m/s)& 1.2e-04(m/s)& 1.2e-04 (m/s)\\
			%			\hline
			$v$               &  0.0027 (m/s)&0.013 (m/s)& 0.0047(m/s)& 0.15 (m/s)& 0.005(m/s)& 0.28 (m/s)\\
			%			\hline
			$\omega_z$                 &6.2e-04 (rad/s)&0.004 (rad/s)&9.01e-04 (rad/s)&0.023 (rad/s)&9.05e-04 (rad/s)&0.04 (rad/s) \\ 
			\bottomrule
		\end{tabular}
	\end{center}
\end{table*}

While the results of the estimator are promising, there are two limiting assumptions of the proposed scheme.  The first is that the estimator assumes the terrain is homogeneous.  This is common among the approaches reported in the literature \cite{Iagnemma2004}. However, in reality, terrain parameters may be changing and evaluating the performance of the estimator in this scenario is subject to future work.  The second limiting assumption is that SCM is treated as the ground truth and the surrogate terramechanics model is parameterized accordingly.  The reality may be different than SCM. However, in that case, a similar procedure in developing the surrogate model could potentially be used by replacing the SCM simulations with an experimental single wheel test bed.  Experimental validation of the developed terramechanics model is subject to future work.

\section{Conclusion} \label{sec:Conclusion}
This paper considers AGVs operating on deformable terrains with unknown terrain properties and develops a novel terrain estimation framework towards increasing the terrain-awareness of the AGV. In particular, the novelty of the framework is the development of a new surrogate terramechanics model for SCM and its use in conjunction with a bicycle model in a UKF. The results suggest that this new framework can estimate the dominant terrain parameter, namely the sinkage exponent, with high accuracy and high computational efficiency. It is therefore concluded that the framework is an important step towards achieving a good balance between estimation accuracy and computational speed. The results also show that the increase in the accuracy of the terrain parameter due to the developed estimation framework leads to a significant increase in the predictive accuracy of the bicycle model, especially for longer prediction time horizons. It is therefore concluded that the proposed framework could be useful to increase the performance of AGVs when they are controlled with model predictive schemes.

Future work includes evaluating the estimator on varying terrain conditions. The need for and ability of estimating multiple terrain parameters also needs to be investigated. It is also of interest to perform experimental validation of the estimator on an actual AGV, and investigate the utility of the estimator in a model predictive control framework for terrain-aware autonomous navigation.

\section*{Appendix}
The following depicts the formulas for $g_1-g_3$ for a clay terrain.  In this work the equations for $g_1-g_3$ are determined for four separate slip ranges for better agreement.  The slip ranges are:
\begin{equation}
0.16\le s \label{slip_range1}
\end{equation}
\begin{equation}
0\le s <0.16 \label{slip_range2}
\end{equation}
\begin{equation}
-0.157< s <0 \label{slip_range3}
\end{equation}
\begin{equation}
s \le-0.157 \label{slip_range4}
\end{equation}
The dependencies for $g_1-g_3$ on each input is determined through curve fitting to simulation data as discussed in Sec. \ref{surrogate}.  The below equations are valid for the slip range of $0\le s <0.16$.
For the lower curve of Fig. \ref{fig:3}, the equations are given as:
\begin{equation}
g_1 =  g_{1_n} g_{1_{s}}\label{g1}
\end{equation}
\begin{equation}
g_2 = g_{2_n} g_{2_{s}} g_{2_{F_z}}  g_{2_v} \label{g2}
\end{equation}
\begin{equation}
g_3 = g_{3_n} g_{3_{s}} g_{3_{F_z}}  \label{g2}
\end{equation}
where
\begin{equation}
\begin{split}
g_{1_n} = 128.3 n^5-415.4 n^4+523.3 n^3\\
-320 n^2+95.07 n-9.942
\end{split}
 \label{scale_g1n}
\end{equation}
\begin{equation}
\begin{split}
g_{1_{s}} = 563.5 s^4-107.9 s^3-4.848 s^2\\
+1.761 s+1.024
\end{split}
 \label{scale_g1k}
\end{equation}
\begin{equation}
g_{2_n} = max(1.8 n-1.08,0) \label{scale_g2n}
\end{equation}
\begin{equation}
\begin{split}
g_{2_{s}} = 1.367\times10^6 s^4-3.71\times10^5 s^3\\
+2.809\times10^4 s^2-545.7 s+70 
\end{split}
\label{scale_g2sll}
\end{equation}
\begin{equation}
g_{2_{F_z}} = (0.0001 F_z+0.7) \label{scale_g2Fz}
\end{equation}
\begin{equation}
g_{2_v} = (4.908 v^{-0.9295}) \label{scale_g2v}
\end{equation}
\begin{equation}
g_{3_n} = -0.1235 n^2+0.7287 n+0.08425 \label{scale_g3n}
\end{equation}
\begin{equation}
\begin{split}
g_{3_{s}} = -309.5 s^4+90.48 s^3-8.983 s^2\\
+0.2631 s+0.086
\end{split}
 \label{scale_g3sl}
\end{equation}
\begin{equation}
g_{3_{F_z}} = (8.3\times10^{-5}) F_z+0.76 \label{scale_g3Fz}
\end{equation}

The same process yields the following equations for the upper curve of Fig. \ref{fig:3}
\begin{equation}
g_1 =  g_{1_{s}}\label{g1u}
\end{equation}
\begin{equation}
g_2 = g_{2_n} g_{2_{s}} g_{2_{F_z}} g_{2_v} \label{g2u}
\end{equation}
\begin{equation}
g_3 = g_{3_n} g_{3_{s}} g_{3_{F_z}} \label{g3u}
\end{equation}
where
\begin{equation}
\begin{split}
g_{1_{s}} = 1.16 (1913 s^4-520.6 s^3+49.57 s^2\\
-1.204 s+1.024)
\end{split}
 \label{scale_g1ku}
\end{equation}
\begin{equation}
g_{2_n} = max(1.8 n-1.08, 0) \label{scale_g2nu}
\end{equation}
\begin{equation}
\begin{split}
g_{2_{s}} = 1.367\times10^6 s^4-3.71\times10^5 s^3+\\
2.809\times10^4 s^2-545.7 s+70 
\end{split}
\label{scale_g2slu}
\end{equation}
\begin{equation}
g_{2_{F_z}} = (0.0001 F_z+0.7) \label{scale_g2Fzu}
\end{equation}
\begin{equation}
g_{2_v} = (4.908 v^{-0.9295}) \label{scale_g2vu}
\end{equation}
\begin{equation}
g_{3_n} = -0.22 \label{scale_g3nu}
\end{equation}
\begin{equation}
g_{3_{s}} = 1609 s^4-529 s^3+58.88 s^2-2.467 s+0.082 \label{scale_g3slu}
\end{equation}
\begin{equation}
g_{3_{F_z}} = (8.3\times10^{-5}) F_z+0.76 \label{scale_g3Fzu}
\end{equation}

The equations for other slip ranges can easily be determined by repeating the curve fitting process on data in those ranges.  Furthermore, it should be noted that these particular equations are only valid for the specific tire under consideration.  Should these be used for a different tire, for example of different radius,  the equations are no longer valid and the process would need to be repeated.

\bibliographystyle{IEEEtran}
\bibliography{terrain_est_publications,additional_refs}

% Generated by IEEEtran.bst, version: 1.14 (2015/08/26)
\begin{thebibliography}{10}
\providecommand{\url}[1]{#1}
\csname url@samestyle\endcsname
\providecommand{\newblock}{\relax}
\providecommand{\bibinfo}[2]{#2}
\providecommand{\BIBentrySTDinterwordspacing}{\spaceskip=0pt\relax}
\providecommand{\BIBentryALTinterwordstretchfactor}{4}
\providecommand{\BIBentryALTinterwordspacing}{\spaceskip=\fontdimen2\font plus
\BIBentryALTinterwordstretchfactor\fontdimen3\font minus
  \fontdimen4\font\relax}
\providecommand{\BIBforeignlanguage}[2]{{%
\expandafter\ifx\csname l@#1\endcsname\relax
\typeout{** WARNING: IEEEtran.bst: No hyphenation pattern has been}%
\typeout{** loaded for the language `#1'. Using the pattern for}%
\typeout{** the default language instead.}%
\else
\language=\csname l@#1\endcsname
\fi
#2}}
\providecommand{\BIBdecl}{\relax}
\BIBdecl

\bibitem{Iagnemma2002}
K.~D. Iagnemma and S.~Dubowsky, ``{Terrain estimation for high-speed
  rough-terrain autonomous vehicle navigation},'' in \emph{Unmanned Ground
  Vehicle Technology IV}, vol. 4715, 2002, pp. 256--266.

\bibitem{Taheri2015}
S.~Taheri, C.~Sandu, S.~Taheri, E.~Pinto, and D.~Gorsich, ``{A technical survey
  on terramechanics models for tire–terrain interaction used in modeling and
  simulation of wheeled vehicles},'' \emph{Journal of Terramechanics}, vol.~57,
  pp. 1--22, 2015.

\bibitem{liu2017}
J.~Liu, P.~Jayakumar, J.~L. Stein, and T.~Ersal, ``Combined speed and steering
  control in high speed autonomous ground vehicles for obstacle avoidance using
  model predictive control,'' \emph{IEEE Transactions on Vehicular Technology},
  vol.~66, no.~10, pp. 8746--8763, 2017.

\bibitem{Liu2018}
J.~Liu, P.~Jayakumar, J.~L. Stein, and T.~Ersal, ``{A nonlinear model
  predictive control formulation for obstacle avoidance in high-speed
  autonomous ground vehicles in unstructured environments},'' \emph{Vehicle
  System Dynamics}, vol.~56, no.~6, pp. 853--882, 2018.

\bibitem{Gallina2014}
A.~Gallina, R.~Krenn, M.~Scharringhausen, T.~Uhl, and B.~Sch{\"{a}}fer,
  ``{Parameter Identification of a Planetary Rover Wheel-Soil Contact Model via
  a Bayesian Approach},'' \emph{Journal of Field Robotics}, vol.~31, no.~1, pp.
  161--175, 2014.

\bibitem{Ishigami2007}
G.~Ishigami, A.~Miwa, K.~Nagatani, and K.~Yoshida, ``{Terramechanics-based
  model for steering maneuver of planetary exploration rovers on loose soil},''
  \emph{Journal of Field Robotics}, vol.~24, no.~3, pp. 233--250, 2007.

\bibitem{Smith2014}
W.~C. Smith, ``{Modeling of Wheel-Soil Interaction for Small Ground Vehicles
  Operating on Granular Soil.}'' Ph.D. dissertation, University of Michigan,
  2014.

\bibitem{Guo2016}
T.~Guo, ``{Power Consumption Models for Tracked and Wheeled Small Unmanned
  Ground Vehicles on Deformable Terrains.}'' Ph.D. dissertation, University of
  Michigan, 2016.

\bibitem{Howard2006}
K.~Iagnemma, ``{Terrain Estimation Methods For Enhanced Autonomous Rover
  Mobility},'' in \emph{Intelligence for Space Robotics}, A.~Howard and
  E.~Tunstel, Eds.\hskip 1em plus 0.5em minus 0.4em\relax TSI Press, 2006,
  ch.~17, p. 425.

\bibitem{Gallina2016}
A.~Gallina, R.~Krenn, and B.~Sch{\"{a}}fer, ``{On the treatment of soft soil
  parameter uncertainties in planetary rover mobility simulations},''
  \emph{Journal of Terramechanics}, vol.~63, pp. 33--47, 2016.

\bibitem{Iagnemma2004}
K.~Iagnemma, S.~Kang, H.~Shibly, and S.~Dubowsky, ``{Online Terrain Parameter
  Estimation for Wheeled Mobile Robots With Application to Planetary Rovers},''
  \emph{IEEE Transactions on Robotics}, vol.~20, no.~5, pp. 921--927, 2004.

\bibitem{ZhenzhongJia2011}
{Zhenzhong Jia}, W.~Smith, and {Huei Peng}, ``{Fast computation of wheel-soil
  interactions for safe and efficient operation of mobile robots},'' in
  \emph{IEEE/RSJ International Conference on Intelligent Robots and Systems},
  2011, pp. 3004--3010.

\bibitem{Krenn2011}
R.~Krenn and A.~Gibbesch, ``Soft soil contact modeling technique for multi-body
  system simulation,'' in \emph{Trends in Computational Contact Mechanics},
  G.~Zavarise and P.~Wriggers, Eds.\hskip 1em plus 0.5em minus 0.4em\relax
  Springer, Berlin, Heidelberg, 2011, pp. 135--155.

\bibitem{Krenn2009}
R.~Krenn and G.~Hirzinger, ``{SCM-a soil contact model for multi-body system
  simulations},'' in \emph{European Regional Conference of the International
  Society for Terrain-Vehicle Systems}, Bremen, 2009.

\bibitem{Bekker1962}
\BIBentryALTinterwordspacing
M.~G. Bekker, \emph{Theory of Land Locomotion}.\hskip 1em plus 0.5em minus
  0.4em\relax Ann Arbor, MI: The University of Michigan Press, 1962. [Online].
  Available: \url{http://hdl.handle.net/2027/mdp.39015000986904}
\BIBentrySTDinterwordspacing

\bibitem{Janosi1961}
Z.~Janosi, B.~Hanamoto, and Ferraris, ``{The analytical determination of
  drawbar pull as a function of slip for tracked vehicles in deformable
  soils},'' in \emph{International Conference of the Mechanics of Soil-Vehicle
  Systems}, Torino, Italy, 1961.

\bibitem{Wong1967}
J.-Y. Wong and A.~Reece, ``{Prediction of rigid wheel performance based on the
  analysis of soil-wheel stresses part I. Performance of driven rigid
  wheels},'' \emph{Journal of Terramechanics}, vol.~4, no.~1, pp. 81--98, 1967.

\bibitem{Chrono}
A.~Tasora, R.~Serban, H.~Mazhar, A.~Pazouki, D.~Melanz, J.~Fleischmann,
  M.~Taylor, H.~Sugiyama, and D.~Negrut, ``{Chrono: An Open Source
  Multi-physics Dynamics Engine},'' in \emph{International Conference on High
  Performance Computing in Science and Engineering}, 2016, pp. 19--49.

\bibitem{Ishigami2008}
G.~Ishigami, ``{Terramechanics-based Analysis and Control for Lunar/Planetary
  Exploration Robots},'' Ph.D. dissertation, Tohoku University, 2008.

\bibitem{Ryu2002}
J.~Ryu, E.~J. Rossetter, and J.~C. Gerdes, ``{Vehicle Sideslip and Roll
  Parameter Estimation using GPS},'' in \emph{International Symposium on
  Advanced Vehicle Control}, Hiroshima, Japan, 2002, pp. 373--380.

\bibitem{Liu2016}
J.~Liu, P.~Jayakumar, J.~L. Stein, and T.~Ersal, ``{A study on model fidelity
  for model predictive control-based obstacle avoidance in high-speed
  autonomous ground vehicles},'' \emph{Vehicle System Dynamics}, vol.~54,
  no.~11, pp. 1629--1650, 2016.

\bibitem{Weiss2008}
C.~Weiss, H.~Tamimi, and A.~Zell, ``{A combination of vision- and
  vibration-based terrain classification},'' in \emph{IEEE/RSJ International
  Conference on Intelligent Robots and Systems}, 2008, pp. 2204--2209.

\bibitem{Wan}
E.~Wan and R.~{Van Der Merwe}, ``{The unscented Kalman filter for nonlinear
  estimation},'' in \emph{IEEE Adaptive Systems for Signal Processing,
  Communications, and Control Symposium}, 2000, pp. 153--158.

\bibitem{Kolas2009}
S.~Kol{\aa}s, B.~Foss, and T.~Schei, ``{Constrained nonlinear state estimation
  based on the UKF approach},'' \emph{Computers {\&} Chemical Engineering},
  vol.~33, no.~8, pp. 1386--1401, 2009.

\bibitem{Wong2001}
J.~Y. Wong, \emph{{Theory of Ground Vehicles}}.\hskip 1em plus 0.5em minus
  0.4em\relax Hoboken, New Jersey: John Wiley, 2001.

\end{thebibliography}

\end{document}